\documentstyle[12pt]{article}
\def\be{\begin{equation}}
\def\ee{\end{equation}}

\newcommand{\suN}{\mbox{SU}(N)}
\newcommand{\suNM}{\mbox{SU}(N+M)}
\newcommand{\bi}{\bibitem}
\newcommand{\eqsys}{(\ref{ydiff})--(\ref{Kdiff})}
\newcommand{\eqsysf}{(\ref{ydiff})--(\ref{Kdiff}), (\ref{constraint})}
\begin{document}
\baselineskip=15.5pt

\begin{titlepage}

\begin{flushright}
hep-th/0104069\\
PUPT-1981\\
\end{flushright}
\vfil

\begin{center}
{\huge Fractional D1-Branes at Finite Temperature}\\
\vspace{3mm}
\end{center}

\vfil
\begin{center}
{\large Christopher P. Herzog and Peter Ouyang}\\
\vspace{1mm}
Joseph Henry Laboratories, Princeton University,\\
Princeton, New Jersey 08544, USA\\
\vspace{3mm}
\end{center}

\vfil

\begin{center}
{\large Abstract}
\end{center}

\noindent
The supergravity dual of $N$ regular and $M$ fractional D1-branes
on the cone over the Einstein manifold $Q^{1,1,1}$ 
has a naked singularity in the infrared.  The supergravity dual of 
$N$ regular and $M$ fractional D3-branes on the conifold also
has such a singularity.  Buchel suggested and Gubser et al. 
have shown that in the D3-brane case, the naked singularity is
cloaked by a horizon at a sufficiently high temperature.
In this paper we derive the system of second-order differential
equations necessary to find such a solution for $Q^{1,1,1}$.  
We also find solutions to this system in perturbation theory that
is valid when the Hawking temperature of the horizon is very high.

\vfil
\begin{flushleft}
April 2001
\end{flushleft}
\vfil
\end{titlepage}
\newpage
\renewcommand{\baselinestretch}{1.1}  


\section{Introduction}

The AdS/CFT correspondence \cite{jthroat, US, EW} has produced a wealth
of new information about strongly coupled conformal gauge theories.  
Considerable effort has also been invested into extending it to 
non-conformal theories.  One recent development is the fascinating
story that has emerged surrounding a certain four
dimensional ${\mathcal N}=1$
supersymmetric $SU(N)\times SU(N+M)$ gauge theory \cite{KS, GHKT}.
The goal of the present work
is to attempt to retell at least part of the same story 
\cite{GHKT} for a ${\mathcal N}=2$ supersymmetric
$SU(N) \times SU(N) \times SU(N+M)$ gauge theory living
in two dimensions.    

We begin by summarizing the story surrounding the ${\mathcal N}=1$
$SU(N)\times SU(N+M)$ gauge theory.
The theory may be realized by
adding $M$ fractional D3-branes (wrapped D5-branes) to $N$ regular
D3-branes at the apex of the conifold, which is defined by the
constraint $\sum_{i=1}^4 z_i^2 = 0$ in ${\bf C}^4$ \cite{KN}.
For $M=0$, this gauge theory reduces to the superconformal
theory dual to the $AdS_5 \times T^{1,1}$ background of type
IIB string theory \cite{KW, MP}.  

In the supergravity dual, the $M$ fractional branes correspond
to $M$ units of RR 3-form flux through the 3-cycle of the compact
space $T^{1,1}$.  This flux changes the background and introduces
the logarithmic running of $\int_{S^2} B_2$, which is related
to the running of field theoretic couplings \cite{KN}.  In turn,
this running causes the RR 5-form flux, which corresponds to
the number of ordinary D3-branes, to grow logarithmically with
the radius \cite{KT}, due to the equation $dF_5 = H_3 \wedge F_3$.
In \cite{KS}, this behavior was attributed to a cascade of Seiberg
dualities in the dual gauge theory.

While the Klebanov-Tseytlin (KT) solution \cite{KT} is smooth
in the UV (for large $\rho$), it has a naked singularity in the
IR.  Two complementary ways have been found of removing the
singularity, and it is with the removal of the singularity that
the story becomes very interesting.  In \cite{KS}, Klebanov
and Strassler (KS) proposed to deform the conifold, i.e. to replace
the constraint with $\sum_{i=1}^4 z_i^2 = \epsilon^2$.
The resulting solution, a warped deformed conifold, is perfectly
non-singular and without a horizon in the IR, while it
asymptotically approaches the KT solution \cite{KT} in the UV.
The mechanism that removes the naked singularity is related
to the breaking of the chiral symmetry in the dual
$SU(N)\times SU(N+M)$ gauge theory.  The ${\bf Z}_{2M}$ chiral
symmetry, which may be approximated by $U(1)$ for large $M$, 
is realized geometrically as $z_i \rightarrow z_i e^{i\theta}$.
The deformation of the conifold breaks it down to 
${\bf Z}_2: \; z_i \rightarrow -z_i$ \cite{KS} although
supersymmetry is preserved.  

The second mechanism for removing the singularity from the KT
solution was proposed by Buchel \cite{Buchel} and later
worked out in detail \cite{BHKPT, GHKT}.  It was suggested
that a non-extremal and
hence supersymmetry breaking generalization of the KT solution, with
unbroken $U(1)$ symmetry, may have a regular Schwarzschild horizon
``cloaking'' the naked singularity.  The dual field theory
interpretation is restoration of chiral symmetry above some
critical temperature $T_c$.  In \cite{BHKPT, GHKT}, the authors
were able to show that at least at high temperatures, where
the differential equations could be analyzed through perturbation
theory in the number of fractional D3-branes, a well behaved supergravity
solution exists with restored $U(1)$ symmetry which in the IR involves
a regular Schwarzschild horizon but which in the UV approaches
the asymptotic KT geometry.  

These two methods of removing the singularity from the KT solution
form an attractive and consistent picture for 
the gauge theory dual.  At high temperature,
above $T_c$, the chiral $U(1)$ symmetry is present.  As we lower
the temperature, the horizon distance shrinks until we reach the critical
temperature $T_c$ where the horizon can no longer ``shield''
the singularity.  At this point, a phase transition occurs and a
KS type solution \cite{KS} becomes preferred.  The chiral symmetry
is broken.   

In \cite{KH}, generalizations of the KT solution were found involving
fractional M2-branes and fractional D$p$-branes, $p=0,1,2,4$.  Like
in the KT solution, the transverse space is conical and moreover these
generalizations are typically smooth in the UV but possess a naked
singularity as the radius of the cone shrinks below some critical
value.  This singularity renders the dual gauge theory poorly defined
in the IR.  A detailed understanding of the gauge/gravity
correspondence for these fractional branes could shed light on gauge
theories in dimensions other than four, and so removing the IR
singularities of these generalizations is an important challenge.  In
the case of fractional D2-branes, a KS type \cite{KS} solution has
already been found \cite{Cvetic3} that resolves the singularity while
preserving supersymmetry.  The IR limit $\rho\rightarrow 0$ of this
fractional D2-brane solution is thought to correspond to confinement
in the dual three dimensional gauge theory.  

It is natural to wonder if the mechanism for singularity resolution at
high Hawking temperature in the fractional D3-brane system will work
for fractional D$p$-branes with $p$ other than 3.\footnote{A related
finite temperature solution was discussed in \cite{Buchel2}, which
used the wrapped brane solution of \cite{MN} to identify appropriate
boundary conditions in the IR.}  We shall focus on the case $p=1$, the
fractional D-string solution.  This solution consists of a warped
product of ${\bf R}^{1,1}$ flat space-time directions and a Ricci
flat, eight dimensional cone ${\bf Y}_8$.  The base of the cone is
then a seven dimensional Einstein space, which in this paper will be
$Q^{1,1,1}$ \cite{classification}.  The manifold $Q^{1,1,1}$ can be
described as a coset:
\[
Q^{1,1,1} = \frac{SU(2)^3}{U(1)^2} \ .
\]
The ordinary
D-strings fill the space-time dimensions.  D3-branes are wrapped over
one of the 2-cycles of $Q^{1,1,1}$, and the remaining two dimensions
of these D3-branes fill ${\bf R}^{1,1}$, creating the fractional
D-strings.  In the dual supergravity description, the fractional
D-strings correspond to turning on an RR five form flux, $F_5$, which
pierces one of the five-cycles of $Q^{1,1,1}$.  Meanwhile, the
ordinary D-strings correspond to an electric form flux $F_3$ in the
${\bf R}^{1,1}$ and radial directions.

Our non-extremal generalization of the fractional D1-brane solution is
well-behaved and singularity free at high Hawking temperature.
Moreover, our ansatz preserves the $U(1)$ symmetry of $Q^{1,1,1}$.
Hopefully, a $U(1)$ breaking deformation of the cone over $Q^{1,1,1}$,
which eliminates the singularity of the KT type fractional D-string
solution, will someday be found.  It seems reasonable to conjecture
that there is some chiral symmetry breaking phase transition, just as
in the fractional D3-brane case \cite{GHKT}. \footnote{Although phase
transitions are usually forbidden for 1+1 dimensional systems, they
can occur for gauge theories in the limit of infinite $N$ \cite{GW}.
This phase transition is a large $N$ phenomenon; for finite $N$ we
would expect a smooth crossover between the symmetric high-temperature
phase and the low-temperature phase.}

Below the critical temperature $T_c$, this hypothetical deformed cone
over $Q^{1,1,1}$ would be preferred.  Above $T_c$, the ansatz we
present below describes a fractional D-string solution where the naked
singularity is cloaked and the $U(1)$ symmetry restored.

We begin by presenting our ansatz for this $U(1)$ symmetry preserving
fractional D-string.  Next, we develop a perturbation theory that is
valid at high temperature, and thereby show that at sufficiently high
temperature, the naked singularity of the corresponding KT type
fractional D-string is ``shielded'' by an event horizon.

\section{Non-Extremal Generalization of the Fractional D1-Brane Ansatz}

We start with an ansatz for the non-extremal fractional D1-branes.  Our strategy 
is similar to that used in finding the non-extremal generalization
of the fractional D3-branes \cite{BHKPT}.  We will add additional warping functions
to the metric that preserve the 
underlying $U(1) \times SU(2)^3$ symmetry
of $Q^{1,1,1}$.  At the same time, we will leave unchanged the geometric 
dependence of the RR and NSNS field strengths on the two- and five-cycles of 
$Q^{1,1,1}$.

The general ansatz for a 10-d Einstein-frame metric consistent with the underlying
symmetries of $Q^{1,1,1}$ involves five functions $x$, $y$, $z$, $w_1$, and $w_2$ 
of a radial coordinate $u$
\be
ds_{10E}^2 = e^{3z} ( e^{-2x} dX_0^2 + e^{2x} dX_1^2 ) + e^{-z} ds_8^2
\label{metric}
\ee
where
\be
ds_8^2  = e^{14y} du^2 + e^{2y} (dM_7)^2 \ , 
\ee
\be
(dM_7)^2 = e^{-12w_1} g_{\psi}^2 + 
e^{2w_1-2w_2} \sum_{i=1}^2 (g_{\theta_i}^2 + g_{\phi_i}^2) +
e^{2w_1+4w_2} ( g_{\theta_3}^2 + g_{\phi_3}^2 )
\ee
and
\be 
g_\psi = \frac{1}{4} (d\psi + \sum_{i=1}^3 \cos\theta_i d\phi_i) \ , \ \ \
g_{\theta_i} = \frac{1}{2\sqrt{2}} d\theta_i \ , \ \ \
g_{\phi_i} = \frac{1}{2\sqrt{2}} \sin\theta_i d\phi_i \ .
\ee
Here $X_0$ is the Euclidean time and $X_1$ is the longitudinal D-string direction.

This metric can be brought into a more familiar D-string form
\be
ds_{10E}^2 = h(\rho)^{-3/4} [ A(\rho) dX_0^2 + dX_1^2 ] +
h(\rho)^{1/4} \left[ \frac{d\rho^2}{B(\rho)} + \rho^2 (dM_7)^2 \right]
\label{normmetric}
\ee
with the redefinitions
\be
h=e^{-4z-8x/3} \ , \ \ \ \rho = e^{y+x/3} \ , \ \ \ A = e^{-4x} \ , \ \ \
e^{14y+2x/3} du^2 = B^{-1} d\rho^2 \ .
\ee
When $w_1=w_2=0$ and $e^{6y} = \rho^6 = \frac{1}{6u}$, the transverse 8-d space
is the cone over $Q^{1,1,1} = M_7$.  Small $u$ thus corresponds to large distances
(where we shall assume that $h$, $A$, $B \rightarrow 1$, and $\rho \rightarrow \infty$)
and vice versa.

The function $w_1$ squashes the $U(1)$ fiber of $Q^{1,1,1}$ relative to
the three two spheres while the function $w_2$ squashes one of the two
spheres relative to the other two.  For comparison, the non-extremal
generalization of the KT solution, which involved the 5-d Einstein
manifold $T^{1,1}$, made use of only one squashing function $w$, not
two.  This $w$, which squashed the $U(1)$ fiber of $T^{1,1}$ relative to
rest of the manifold, roughly corresponds to our $w_1$.  
The most general volume preserving squashing of $Q^{1,1,1}$ 
consistent with the symmetries would involve
two more functions $w_2$ and $w_3$. 
Note, however, from (\ref{nsnsb}), that the harmonic two-forms on 
$Q^{1,1,1}$ involve linear combinations of the $SU(2)$'s.  In order
to keep one of these harmonic two-forms $u$ independent, we
may only add one more squashing function $w_2$. 
The additional Einstein
equations in the $Q^{1,1,1}$ directions give additional constraints
when compared to the five dimensional $T^{1,1}$ case;
fortunately $w_1$ and $w_2$ give us the freedom necessary to satisfy the
constraints.  Ricci-flat 8-d spaces with nontrivial $w_i$ correspond
to the resolutions of $Q^{1,1,1}$ considered by \cite{Cvetic2}.

The extremal D-string solution and the more general fractional D-string solution
on ${\bf Y}_8$ have $x=w_i=0$.  Adding a non-constant $x(u)$ drives the non-extremality.
For example, the non-extremal version of a D-string on the cone over $Q^{1,1,1}$ 
($w=0$) has $x=au$, $e^{-4x}=A=1-\frac{2a}{3\rho^6}$, 
$e^{-4z-8x/3}=h=1+\frac{\tilde q}{\rho^6}$, $\rho = e^{y+x/3}$.
Our aim will be to understand how switching on the non-extremality ($x=au$)
changes the extremal fractional D-string solution.

Our ansatz for the $p$-form fields is dictated by the geometry and thus
is exactly the same as in the extremal fractional D-string case \cite{KH}:
\be
F_3 = K(u) e^{6z-\Phi} d^2x \wedge du \ ,
\ee
\be
B_2 = f(u) \omega_2 \ , \ \ \ 
\omega_2 = (g_{\theta_1} \wedge g_{\phi_1} - g_{\theta_2} \wedge g_{\phi_2}) / \sqrt{2} \ ,
\label{nsnsb}
\ee
\be
F_5 = P (\omega_5 + {*}\omega_5) \ , \ \ \ 
\omega_5 = -g_{\psi} \wedge \omega_2 \wedge g_{\theta_3} \wedge g_{\phi_3} \ .
\ee
Moreover, the dilaton $\Phi$ is assumed to be a function of the radial 
variable $u$ only.
The $M$ fractional D-strings (wrapped D3-branes) thus correspond 
$M$ units of flux through the five-cycle of $Q^{1,1,1}$,
and $P \sim g_s M$.  The function
$K(u)$ corresponds roughly to the flux of ordinary D-strings through 
the compact space $Q^{1,1,1}$ itself.
The equation of motion for $F_3$, $d{*}e^{\Phi} F_3 = F_5 \wedge H_3$, implies
\be 
K(u) = Q + P f(u) \ .
\label{intK}
\ee
In what follows, we shall use this ansatz to reduce the type IIB supergravity 
equations of motion to a system of nonlinear, coupled ordinary differential
equations describing the radial evolution of $x$, $y$, $z$, $w_1$, $w_2$, $K$, and
$\Phi$.

\section{Derivation of the Equations of Motion}

We have seven warping functions and hence will require a system of seven
ordinary differential equations.  From analogy with the non-extremal
generalization of the KT solution \cite{BHKPT}, we also expect a zero energy constraint, giving eight equations total.
Consideration of the $p$-form field strengths yields two nontrivial 
equations of motion, one for the dilaton and one for $H_3 = dB_2$.  The $H_3$
equation of motion, $d{*}e^{-\Phi} H_3 = -F_5 \wedge F_3$, reduces to the
ordinary differential equation
\be
(e^{2z-4y-4w_1+4w_2-\Phi} K')' = P^2 K e^{6z-\Phi}\ ,
\ee
while the dilaton equation of motion, 
$2 \, d{*}d\Phi = -e^{-\Phi} H_3 \wedge {*} H_3 + e^{\Phi} F_3 \wedge {*}F_3$,
reduces to
\be
2 \Phi'' = -e^{-\Phi+2z-4y-4w_1+4w_2} \frac{K'^2}{P^2} - K^2 e^{6z-\Phi} \ .
\label{phiprelim}
\ee

Einstein's equations are $R_{MN} = T_{MN}$ where $R_{MN}$ is the
Ricci curvature and 
\begin{eqnarray}
 T_{MN} &=& \frac{1}{2} \partial_M \Phi \partial_N \Phi +
\frac{1}{96}
\tilde F_{MPQRS} \tilde F_N{}^{PQRS}
\nonumber\\
&& +\frac{1}{4}(e^{-\Phi} H_{MPQ} H_N{}^{PQ} +
e^{\Phi} \tilde F_{MPQ} \tilde F_N{}^{PQ}) \nonumber\\
&&
- \frac{1}{48} G_{MN} (e^{-\Phi} H_{PQR} H^{PQR} +
 e^{\Phi} \tilde F_{PQR} \tilde F^{PQR})\ . 
\label{Einstein}
\end{eqnarray}
In order to write down these equations in a convenient form, 
we will work in an orthonormal frame basis:
\be
e^{0} = e^{\frac{3}{2}z-x} dX_0 \ , \ \ \ \
e^{1} = e^{\frac{3}{2}z+x} dX_1 \ , \ \ \ \
e^{u} = e^{-\frac{1}{2}z+7y} du \ ,
\ee
\be
e^{\psi} = e^{-\frac{1}{2}z +y -6w_1} g_\psi \ , \ \ \ \
e^{\theta_3} = e^{-\frac{1}{2}z+y+w_1+2w_2}g_{\theta_3} \ , \ \ \ \
e^{\phi_3} = e^{-\frac{1}{2}z+y+w_1+2w_2}g_{\phi_3} \ ,
\ee
\be
e^{\theta_i} = e^{-\frac{1}{2}z+y+w_1-w_2}g_{\theta_i} \ , \ \ \ \
e^{\phi_i} = e^{-\frac{1}{2}z+y+w_1-w_2}g_{\phi_i} \ , \ \ \ i=1,2  \ . 
\ee
In this basis, Einstein's equations are diagonal.  Moreover, the
equations corresponding to 
$R_{\theta_1\theta_1}$, $R_{\theta_2\theta_2}$, $R_{\phi_1\phi_1}$, and $R_{\phi_2\phi_2}$
are identical and similarly for the equations corresponding to $R_{\theta_3\theta_3}$
and $R_{\phi_3\phi_3}$.  Thus, we are left with six relations.
Our strategy will
be to put aside $R_{uu}$ at first and use the remaining five relations along
with the two field strength equations to derive a 
second order, nonlinear system in the seven warping functions.  At the end, we
will find that the $R_{uu}$ relation provides a zero energy constraint, analogous
to the one found in \cite{BHKPT}.

{}From the field strengths, it is relatively easy to see that $T_{00}=T_{11}$.  However,
$R_{00}=-e^{z-14y}(\frac{3}{2}z'' + x'')$, while $R_{11}=-e^{z-14y}(\frac{3}{2}z'' - x'')$.
Hence, the first two Einstein equations allow us to solve for $x(u)$ exactly:
\be
x''=0 \ , \ \ \ x=au \ , \ \ \ a>0 \ .
\label{xsol}
\ee
In the case of non-extremal fractional D3-branes \cite{BHKPT}, 
the same behavior was found for this $x$ function, 
and the factor $a$ was identified
with the degree of non-extremality.

Having solved for $x$, we can use either of the first two Einstein
equations to find an equation for $z$:
\be
12 z'' = 2P^2 e^{4z+4y+4w_1-4w_2} +3K^2 e^{6z-\Phi} + 
\frac{(K')^2}{P^2}e^{2z-4y-4w_1+4w_2-\Phi} .
\label{zprelim}
\ee
Comparing (\ref{phiprelim}) and (\ref{zprelim}) with the corresponding
equations (3.18) and (3.19) in \cite{BHKPT}, there is some interesting similarity.  Both
here and in \cite{BHKPT}, the equations involve sums of the same three
terms proportional to $P^2$, $K^2$, and $(K'/P)^2$.  In fact, there exists
a linear transformation of (\ref{phiprelim}) and (\ref{zprelim}),
\[
z = z_n + \frac{1}{4} \Phi_n \ , \ \ \ \ 
z_n = \frac{3}{4} z - \frac{1}{8} \Phi \ ,
\]
\be
\Phi = \frac{3}{2} \Phi_n - 2 z_n \ , \ \ \ \
\Phi_n =  \frac{1}{2} \Phi + z \ , 
\label{zptrans}
\ee
such that the new differential equations
(\ref{phin}) and (\ref{zn})
correspond almost precisely with those in \cite{BHKPT}.  To take
advantage of the calculations in \cite{BHKPT} and \cite{GHKT}, we
shall use the transformed variables $z_n$ and $\Phi_n$ in what
follows.  Note that if $\Phi_n'' = 0$, then $z'' = -\Phi''/2$.
Moreover, in the extremal case $x=0$, $z=-\Phi/2$ and
$h^{-3/4}=e^{3z}=h^{-1/2}e^{-\Phi/2}$.  Hence, we see that the
Einstein frame metric (\ref{normmetric}) in this particular case can
be obtained from the string frame metric through the usual procedure
of multiplying by a factor of $e^{-\Phi/2}$.

To derive similar equations for $y$, $w_1$, and $w_2$, we need to take linear
combinations of the Einstein equations involving 
\begin{eqnarray*}
R_{\psi\psi} &=& e^{z-14y} ( \frac{1}{2}z'' - y'' + 6w_1'')
+ 2e^{z-2y}(e^{-16w_1-8w_2} + 2e^{-16w_1+4w_2}) \ , \\
R_{\theta_1\theta_1} &=& e^{z-14y} ( \frac{1}{2}z'' - y'' - w_1'' + w_2'')
+2e^{z-2y}(-e^{-16w_1+4w_2}+4e^{-2w_1+2w_2}) \ , \\
R_{\theta_3\theta_3} &=& e^{z-14y} ( \frac{1}{2}z'' - y'' - w_1'' - 2w_2'')
+2e^{z-2y}(-e^{-16w_1-8w_2}+4e^{-2w_1-4w_2}) \ .
\end{eqnarray*}
We leave it to the reader to derive the field strength contributions (\ref{Einstein})
to Einstein's equations and give only the results
\be
7y'' = -\Phi_n'' + 2e^{12y} ( -e^{-16w_1-8w_2} -2e^{-16w_1 + 4w_2}  + 
8e^{-2w_1-4w_2} + 16e^{-2w_1+2w_2}) \ ,
\label{ydiff}
\ee
\be
7w_1'' = \Phi_n'' + \frac{8}{3} e^{12y}(-e^{-16w_1-8w_2}-2e^{-16w_1+4w_2}
+ e^{-2w_1-4w_2} + 2e^{-2w_1+2w_2}) \ ,
\label{wadiff}
\ee
\be
2w_2'' = -\Phi_n'' - \frac{4}{3} e^{12y}( e^{-16w_1-8w_2} - e^{-16w_1+4w_2}
- 4e^{-2w_1-4w_2} + 4e^{-2w_1+2w_2}) \ ,
\label{wbdiff}
\ee
\be
6 \Phi_n'' = P^2 e^{4z_n + 4y+\Phi_n + 4w_1-4w_2}
-\frac{(K')^2}{P^2} e^{4z_n-4y-\Phi_n-4w_1+4w_2} \ ,
\label{phin}
\ee
\be
8z_n'' = 2K^2 e^{8z_n} + P^2 e^{4z_n + 4y+\Phi_n + 4w_1-4w_2}
+\frac{(K')^2}{P^2} e^{4z_n-4y-\Phi_n-4w_1+4w_2} \ ,
\label{zn}
\ee
\be
(e^{4z_n-4y-\Phi_n - 4w_1+4w_2} K')' = P^2 K e^{8z_n} \ .
\label{Kdiff}
\ee
As yet, we have ignored the $R_{uu}$ Einstein equation.  For our metric
\be
R_{uu} = e^{z-14y} \left( \frac{1}{2} z''-6(z')^2 - 7y'' +42(y')^2
-2(x')^2 -42(w_1')^2 -12(w_2')^2 \right)
\ee
which, using (\ref{ydiff}), (\ref{zn}), (\ref{phin}),
(\ref{zptrans}) and (\ref{xsol}), yields the zero energy constraint
\[
84(y')^2 -16(z_n')^2 -3(\Phi_n')^2 -84(w_1')^2-24(w_2')^2
+6\Phi_n'' + K^2 e^{8z_n} - 
\]
\be
4e^{12y}(-e^{-16w_1-8w_2} -2e^{-16w_1 + 4w_2}  + 
8e^{-2w_1-4w_2} + 16 e^{-2w_1+2w_2}) = 4a^2 \ .
\label{constraint}
\ee
In later sections, it will be important to keep track of the
dimensions of the various parameters involved.  From the form of the
metric (\ref{metric}) it is natural to require that $e^y$ and
$u^{-1/6}$ should have dimension of length, while $x,z,w$ be
dimensionless.  Since we have set the 10-d gravitational constant
to be 1 (i.e.~we measure the scales in terms of the 10-d ``Planck
scale'' $L_{\rm P} \sim (g_s \alpha'^2)^{1/4}$), then from
(\ref{constraint}), we conclude that $K$ and $Q$ in ({\ref{intK}) have
dimension (length$)^6$ while $P$ has dimension (length$)^4$ and $f$
has dimension (length$)^2$.  It is easy to restore the dependence on
the Planck length by rescaling $Q \rightarrow L_{\rm P}^6 Q$, $P
\rightarrow L_{\rm P}^4 P$, etc.  To restore the dependence on the
string coupling one should further rescale $P^2 \rightarrow g_s P^2$.
At the end, $Q \sim g_s \alpha'^3 N$, $P \sim g_s \alpha'^2 M$, where
$N$ and $M$ are the number of ordinary and fractional D-strings,
respectively.

\section{Three Simple Solutions}

In addition to the extremal D-string solution, there are three other
relatively simple solutions to the system of equations \eqsysf.  Some
of of these solutions were discussed briefly in the Introduction and
in Section 2.  There is the analog of the KT solution for fractional
D-strings, what we will call the extremal fractional D-string
solution.  Next, there is the non-extremal ordinary D-string solution.
These two solutions will be very important when we try to find a
non-extremal fractional D-string solution through perturbing in the
number of $P$ of fractional D-strings.  We will find a solution which
interpolates between the extremal fractional D-string solution in the
UV and the non-extremal ordinary D-string solution in the infrared.

The third solution is more of a mathematical curiosity.  It is the
analog of the singular, non-extremal D3-brane solution found in
\cite{Buchel}.  This third solution, although non-extremal, is
singular because it has a naked singularity in the far infrared.
Although it approaches the extremal fractional D-string solution as
the non-extremality parameter $a \rightarrow 0$, it does not approach
the non-extremal ordinary D-string solution as the number of
fractional D-strings $P \rightarrow 0$.

\subsection{Singular Non-Extremal Fractional D-Strings}

As a first attempt at finding a non-extremal solution, we might be
tempted to try to preserve the geometry of the $Q^{1,1,1}$ base space
by turning off the squashing functions $w_i$; this approach will lead
us to the D-string analog of the Buchel solution.  Our motivation is
two-fold.  First, because this solution is singular, we will see the
necessity of squashing the $Q^{1,1,1}$.  Second, to obtain the
extremal fractional D-strings in the next subsection, we need only
take the limit in which the non-extremality parameter $a \rightarrow
0$.

So let us suppose that the $Q^{1,1,1}$ is not squashed,
i.e. $w_1=w_2=0$.  Then the equations \eqsysf\ simplify dramatically.  From
(\ref{wbdiff}), we have $\Phi_n'=p$, where $p$ is a constant, so (\ref{phin}) becomes
\be 
f' = -P e^{4y + \Phi_n} \ .
\label{fprime}
\ee
{}From (\ref{Kdiff}), it is straightforward to see that $z_n$ must then
also satisfy a first order equation:
\be
(e^{-4z_n})' = Q+Pf \ .
\label{znprime}
\ee
Then (\ref{ydiff}) reduces to $y'' = 6e^{12y}$.
The zero-energy constraint (\ref{constraint}) sets one of the 
integration constants of this differential equation:
\be 
y' = -\sqrt{b^2 + e^{12y}} \ , \ \ \ \ 84 b^2 = 4a^2 + 3p^2 \ ,
\label{xyz}
\ee
which integrates to give $e^{6y} = \frac{b}{\sinh 6bu}$.  Without
loss of generality, we may choose $b>0$.  The 
differential equation for $f$ (\ref{fprime}) becomes
\be
f' = -P e^{(p-4b)u} \left(\frac{2b}{1-e^{-12bu}}\right)^{2/3} \ .
\ee
Once $f$ is known, $z_n$ is easily found by integrating (\ref{znprime}).

The precise analog with the Buchel solution is found by taking $p=-2a/3$ 
which in turn implies that $b=2a/3\sqrt{7}$ (\ref{xyz}), but we shall
keep $p$ and $b$ general in the discussion that follows.
To demonstrate the pathological nature of these solutions, let us take a look
at the Ricci scalar:
\be
R = \frac{1}{4} e^{z_n-14y+pu/4} \left( P^2 e^{4y+4z_n+pu} 
-\frac{1}{2} \left( (Q+Pf)^2e^{8z_n} 
-6p(Q+Pf)e^{4z_n} - 9p^2 \right) \right)
\ee
In analogy with the behavior of the Buchel solution for D3-branes, we
expect this solution will exhibit divergences at large $u$, so we develop
asymptotics
for $f$ and $z_n$ far in the IR:
\be
f = f_0 -\frac{P(2b)^{2/3}}{p-4b}e^{(p-4b)u} + O(e^{(p-16b)u})
\label{aso}
\ee
and hence
\be
e^{-4z_n} = C_1 + u(Q + Pf_0) - 
\frac{P^2(2b)^{2/3}}{(p-4b)^2} e^{(p-4b)u} + O(e^{(p-16b)u})  \ 
\label{ast}
\ee
where $C_1$ is some integration constant.  
The asymptotics are clearly very sensitive to the relative magnitudes
of $p$ and $b$.  Hence we will consider the two cases,
$p-4b \geq 0$ and $p-4b < 0$ separately.

In the case $p-4b \geq 0$, the exponential term in (\ref{ast}) is
dominant and causes $e^{-4z_n}$ to be large and negative at large $u$.
The subsequent exponential terms contribute very little because of the
constraint (\ref{xyz}): in particular, in order for $a$ to be real,
$|p| \leq \sqrt{28} b$ and hence $p-16b < 0$.  The function
$e^{-4z_n}$ is continuous in the region $0<u<\infty$, and note that
near $u=0$, $e^{-4z_n} \sim C_2 + u(Q + Pf_0) + O(u^{4/3})$.  Hence,
close to $u=0$, $e^{-4z_n}$ is nonnegative (in order to have well
behaved asymptotics at large distances) and growing.  From these
facts, it is clear that $e^{-4z_n}$ becomes zero at some finite
$u_{cr}$ and the Ricci scalar blows up.  Defining the horizon as the
locus where $G_{00}=\exp(3z_n-2au+3pu/4)$ vanishes, a horizon will
only occur, if at all, in the limit $u\rightarrow \infty$.  Hence, the
singularity is naked and the solution is pathological.

{}For the case $p-4b<0$, the constant and linear terms in (\ref{aso})
and (\ref{ast}) dominate at large $u$.  Hence, $f \sim f_0$ and
$e^{-4z_n} \sim u$.  Also, $e^{-y} \sim e^{bu}$.  As was noted
previously, in order for $a$ to be real, $\sqrt{28} b \geq |p|$.
Hence, the $e^y$ terms will dominate the Ricci scalar and send it off
to infinity at large $u$, even if we consider the limit $P\rightarrow
0$.  In other words, this solution has a naked singularity in the IR,
even in the absence of fractional D-strings.  The best we can do is to
set the relative magnitudes of the $a$ and $p$ so that there is also a
horizon at $u=\infty$.  These considerations show that leaving the
$Q^{1,1,1}$ unsquashed leads to a naked singularity, so to find a
well behaved non-extremal fractional D-string solution, we will have
to look elsewhere.

\subsection{The Extremal Fractional D-string Solution}

As was mentioned above, the limit $a, \, p \rightarrow 0$ of 
the singular non-extremal D-string solution recovers
the extremal fractional D-string solution.  Let us take this
limit:
\be
\Phi_n \rightarrow 0 \ , \ \ \ \
e^{6y}  \rightarrow \frac{1}{6u} \ , \ \ \ \
f \rightarrow f_0 -  \frac{6^{1/3} P}{2} u^{1/3}  \ , 
\label{extone}
\ee
\be
e^{-4z_n} \rightarrow C_1 + (Q+Pf_0)u - \frac{3 \cdot 6^{1/3} P^2}{8} u^{4/3} \ .
\label{exttwo}
\ee
This solution 
is well
behaved in the UV, in the limit $u \rightarrow 0$.
If $C_1 = 1$, then the metric (\ref{normmetric}) becomes asymptotically
flat at large distance $u \rightarrow 0$.  However, in what follows,
it will be more convenient to choose $C_1=0$, thus eliminating the asymptotically
flat regime, ``zooming in'' on the low-energy dynamics of the dual gauge theory.
As was noted in \cite{KH}, the solution gives rise to a naked singularity in 
the infrared at $u_{cr}^{1/3} = (Q+Pf_0) 4 \cdot 6^{2/3} / 9 P^2$.
In analogy with \cite{BHKPT}, we shall define $L_{\rm P}$ as the value
$u_{cr}^{-1/6}$ at which this solution develops a singularity.

\subsection{The Non-Extremal Ordinary D-String Solution}

We are searching for a solution without fractional D-strings, so we set $f=P=0$.
Moreover, we know from the literature \cite{Peet} that this solution does not 
require the extra degrees of freedom provided by $w_1$ and $w_2$ so we set
$w_1=w_2=0$.  The system of equations \eqsysf\ simplifies 
substantially in this case and becomes easily integrable:
\be
y'' = 6e^{12y} \ , \ \ \ \ z_n'' = \frac{1}{4} Q^2 e^{8z_n} \ , \ \ \ \
x'' = 0 \ , \ \ \ \ \Phi_n'' = 0 \ .
\label{abc}
\ee
Integrating these equations once, we find
\be
x' = a \ , \ \ \ \ \Phi_n' = p \ , \ \ \ \ 
y'^2 = b^2 + e^{12y} \ , \ \ \ \ 
z_n'^2 = c^2 + \frac{Q^2}{16} e^{8z_n} \ . 
\ee
{}From the constraint (\ref{constraint}), we find that
the integration constants must obey $a^2-21b^2+4c^2 + \frac{3}{4}p^2=0$.
Integrating the equations for $y$ and $z_n$, we find that
\be
e^{6y} = \frac{b}{\sinh 6bu} \ , \ \ \ \ 
e^{4z_n} = \frac{4c}{Q \sinh 4c(u+k)} \ .
\label{yznex}
\ee
Cast in a more familiar form, 
\be
\rho^6 = e^{6y+2x} = \frac{2be^{-2(3b-a)u}}{1-e^{-12bu}} \ ,
\label{nerho}
\ee
\be
A(u) = e^{-4x} = e^{-4au} \ ,
\ee
\be
h(u) = e^{-4z_n - \Phi_n - 8x/3} = \frac{Q}{4c} e^{-pu-8au/3} \sinh 4c(u+k) \ .
\label{neh}
\ee
As has become customary in gauge/gravity duality, we set the boundary
condition that $h$ approach zero as 
$u \rightarrow 0$.  In other words $k=0$.  This boundary condition
removes the asymptotically flat region so that we
``zoom in'' on the low-energy dynamics of the dual gauge theory.

To match the standard non-extremal D-string solution, we require that
$3b=2c=a$, and $p=-2a/3$.  This choice satisfies the 
zero-energy constraint and 
(\ref{nerho})--(\ref{neh}) become 
\be
h(\rho) = \frac{Q}{6\rho^6} \ , \ \ \ \ A(\rho)=B(\rho)=1-\frac{2a}{3\rho^6} \ .
\ee
In other words, we recover the standard non-extremal D-string metric.  To summarize, the non-extremal ordinary D-string solution consists of $w_i=0$ and
\be
\Phi_n = -2au/3 \ , \ \ \ \
e^{4x} = e^{4au} \ , \ \ \ \
e^{-6y} = 3 a^{-1} \sinh 2au \ , \ \ \ \
e^{-4z_n} = \frac{Q}{2a} \sinh 2au \ .
\label{nexord}
\ee

\section{Asymptotics of the Regular Non-Extremal Fractional D-String}

We have not succeeded in finding an analytic solution to the system of
differential equations \eqsysf.  To proceed further, we can either integrate
the equations numerically or seek an approximate solution in
perturbation theory.  In any case we must be certain that our solution
satisfies the correct boundary conditions in the short-distance ($u
\rightarrow \infty$) and long-distance ($u \rightarrow 0$) limits,
where we understand the physics.  The procedure we will follow is very
similar to that in \cite{GHKT}.

For $P \rightarrow 0$ we must obtain the black D-string solution, which has a regular Schwarzschild horizon.  If the horizon is preserved as we add fractional D1-branes, we expect the following asymptotics to hold as $u \rightarrow \infty$ (\ref{nexord}):
\[
x = au \ , \ \ \ \ y \rightarrow -au/3 +y_* \ , \ \ \ \ 
z_n \rightarrow -au/2 + z_*\ ,
\]
\be
w_i \rightarrow w_{i*} \ , \ \ \ \ \Phi_n \rightarrow -2au/3 + \Phi_* \ , \ \ \ \
K \rightarrow K_* \ .
\label{asympt}
\ee
The metric for $u \rightarrow \infty$ in the $u-X_0$ directions is given by
\be
ds^2 = e^{-4au+3z_*+\frac{3}{4}\Phi_*} dX_0^2 + e^{-4au-z_*-\frac{1}{4}\Phi_* +14y_*}du^2
\ee
so that with the natural near-horizon variable $U=e^{-2au}$, the usual procedure of choosing periodicity of the Euclidean time $X_0$ to avoid a conical singularity fixes the Hawking temperature $T_H$:

\be
T_H = \frac{a}{\pi} e^{2z_* - 7y_* + \Phi_*/2} \ .
\label{th}
\ee

At large distances ($u\rightarrow 0$) the non-extremal solution should 
approach the extremal solution (\ref{extone}), (\ref{exttwo}), i.e.
we require that
\be
u\rightarrow 0 \ : \ \ \ \ x,w,\Phi_n \rightarrow 0 \ , \ \ \ \
y \rightarrow -\frac{1}{6} \log 6u \ .
\label{bcsmall}
\ee
(Note that this behavior is also in agreement with the small $u$ asymptotics
(\ref{nexord}) of the regular non-extremal D-string solution.)  The behaviors
of the effective D-string charge and of the warp factor at small $u$ are 
\be
K(u) \rightarrow \frac{6^{1/3} P^2}{2} \left( 
\frac{3}{4} L_{\rm P}^{-2} - u^{1/3} \right) \ , \ \ \ \
e^{-4z_n} \rightarrow \frac{3 \cdot 6^{1/3} P^2}{8} u 
( L_{\rm P}^{-2} - u^{1/3}) \ .
\label{smallu}
\ee

We expect that the physics of this fractional D-string system should
be very similar to that of the fractional D3-brane solution considered
in \cite{GHKT}.  On the supergravity side, when $T<T_c$ the solution
which preserves the $U(1)$ symmetry of $Q^{1,1,1}$ is singular and one
needs an appropriate deformation, perhaps of the KS-type \cite{KS},
that breaks this $U(1)$ symmetry and removes the singularity.  For
$T>T_c$ we should be able to construct a nonsingular solution which
preserves the $U(1)$ chiral symmetry, and we will do this in
perturbation theory in the next section.

The flux corresponding to the number of D-strings is given by ${*}F_3
= K(u) e^{-\Phi} \omega_2 \wedge \omega_5$, so on the gauge theory
side, we want to think of $K(u)e^{-\Phi}$, the effective D-string charge, as an
effective number of unconfined color degrees of freedom.  As we run
the scale of the theory into the infrared ($u \rightarrow \infty$), the
number of colors should decrease.  Above the critical temperature,
this number will be positive everywhere, but below the critical
temperature, the effective number of colors will vanish at finite $u$.  This behavior is the same as for fractional D3-branes.  
One potentially bothersome difference from the D3-brane case is the dependence of the 
D-string flux on the dilaton.  However, things are all right because
\be
e^{-\Phi} = e^{-\frac{3}{2} \Phi_n + 2 z_n} \propto \frac{e^{au}}{ \sqrt{\sinh{2a(u+k)}}}
\label{dilatonrun}
\ee 
is decreasing for all positive $u$. Thus if $K(u)$ decreases with
increasing $u$, the flux will still decrease as well, provided that
the fractional D-strings are a small enough perturbation that the
variation of the dilaton is dominated by the presence of the ordinary
D-strings.  Moreover, $K(u)e^{-\Phi}$ is well-defined for all values
of $u$, in particular as $u \rightarrow \infty$.  Notice that in
(\ref{dilatonrun}) we have temporarily restored the integration
constant $k$ from (\ref{yznex}).  In Section 4.3, we set $k=0$ to zoom
in on the IR physics.  This approximation introduces a singularity at
$u=0$ ($e^{-\Phi}$ diverges) but this is strictly
an artifact of having removed the asymptotically flat region. 

While the number of colors decreases as we move toward the horizon,
from (\ref{dilatonrun}) one can see that the string coupling $e^\Phi$
increases.  From \cite{IMSY}, we expect that the string coupling can
be expressed in terms of gauge theory parameters as $e^\Phi \sim
g_{YM}^3 N^{1/2} / \Lambda^3$ where $\Lambda$ sets the energy scale of
the gauge theory, and $N \sim K e^{-\Phi}$ gives the number of
ordinary D-strings at a given scale.  Hence, $e^\Phi \sim g_{YM}^2
K^{1/3} / \Lambda^2$.  In the case where $K$ is constant, i.e. there
are no fractional D-strings, we expect the string coupling to become
larger in the IR.  Indeed, (\ref{dilatonrun}) holds exactly in this
case.  However, once we add fractional D-strings, $K$ should decrease
as $u \rightarrow \infty$.  Thus, in perturbation theory, we expect to
see corrections to (\ref{dilatonrun}) that tend to decrease $e^\Phi$
as $u\rightarrow \infty$.

\section{Perturbation Theory in $P$}

One useful approach to constructing the required non-extremal 
fractional brane solution
is to start with the non-extremal ordinary 
D-string solution (\ref{nexord}), which 
is valid for $P=0$, and find its deformation order by order in $P^2$.  A
remarkable feature of perturbation theory in $P^2$ near the extremal
($a=0$) D-string background is that already the {\it first-order}
correction gives the {\it exact} form of the extremal fractional D-string 
solution (\ref{extone}), (\ref{exttwo}).  Therefore, it is natural to expect
that a similar expansion near the non-extremal D-string solution will
capture the basic features of the non-extremal generalization of the
extremal fractional D-string.

More precisely, our starting point will be the well-known non-extremal
ordinary D-string solution (\ref{nexord}) with $Q$ replaced by the effective
D-string charge $K_*$, so that we automatically match onto the near horizon
asymptotics (\ref{asympt}).  Perturbing in $P^2$ around the non-extremal
D-string solution of charge $K_*$, we will see that the $O(P^2)$ modification
is already enough to match onto the extremal fractional D-string long-distance
asymptotics.  The small parameter governing this expansion is actually the
dimensionless ratio $\lambda \equiv P^2 K_*^{-1} a^{-1/3}$, \footnote{
Note that $P \sim g_s M$ and $K \sim g_s N$ where $M$ and $N$ are the numbers
of fractional and regular D-strings respectively.}
i.e. for this method to work the horizon value of the 
effective D-string charge $K_*$ has to be sufficiently large.  In view
of the discussion in Section 5, this means that this perturbation theory
is applicable for $T \gg T_c$.  
Unlike \cite{GHKT}, 
$\lambda$ here depends also on the non-extremality parameter $a$.

It is useful to rescale the fields by appropriate powers of $P^2$, setting
\be
K(u) = K_* + P^2 F(u) \ , \ \ \ \
\Phi_n(u) = -2au/3 + P^2 \phi(u) \ , \ \ \ \
w_i(u) = P^2 \omega_i(u) \ ,
\label{require}
\ee
and
\be
y \rightarrow y + P^2 \xi \ , \ \ \ \
z_n \rightarrow z_n + P^2 \eta \ ,
\ee
where $y$ and $z$ represent the pure D-string solution (\ref{nexord}):
$e^{-6y} = 3a^{-1} \sinh 2au$, $e^{-4z_n} = \frac{K_*}{2a} \sinh 2au$,
and $\xi$ and $\eta$ are corrections to it.  To match onto the small $u$
extremal fractional D-string asymptotics (\ref{smallu}), we require that
\be
\omega_i(0) = \xi(0) = \phi(0) = 0 \ , \ \ \ 
F \rightarrow -\frac{(6u)^{1/3}}{2} \ , \ \ \ \
\eta \rightarrow \frac{u^{1/3} L_{\rm P}^2}{4} \ .
\label{requirem}
\ee

Now the system \eqsys\ takes the following explicit form:
\be
7\xi'' + \phi'' - 504 e^{12y} \xi +O(P^2) = 0 \ ,
\label{xidiff}
\ee
\be
7\omega_1'' - \phi'' -112 e^{12y} \omega_1 +O(P^2) = 0 \ ,
\label{omegai}
\ee
\be
2\omega_2'' + \phi'' +16 e^{12y} \omega_2 + O(P^2) = 0 \ ,
\label{omegaii}
\ee
\be
6\phi'' + e^{4z_n-4y+2au/3} (F'^2 - e^{8y-4au/3}) + O(P^2) = 0 \ ,
\label{phi}
\ee
\be 
(e^{4z_n-4y+2au/3} F')' - K_* e^{8z_n} + O(P^2) = 0
\label{Fdiff}
\ee
\be
8\eta'' -4K_* e^{8z_n} F - 16 K_*^2 e^{8z_n} \eta 
-e^{4z_n-4y+2au/3}(F'^2 + e^{8y-4au/3}) + O(P^2) = 0 \ .
\label{etadiff}
\ee
The constraint (\ref{constraint}) becomes
\[
168 y' \xi' -32 z'\eta' +4a\phi' - e^{4z_n-4y+2au/3} (F'^2 - e^{8y-4au/3})
\]
\be
+8 K_*^2 e^{8z_n} \eta + 2 K_* e^{8z_n} F - 84\cdot 12 e^{12y}\xi + O(P^2) = 0 \ .
\ee

\subsection{Leading-order solution for $K$}

Using the fact that $K_* e^{8z_n} = 4 K_*^{-1} z_n''$ (\ref{abc}),
we get from (\ref{Fdiff}) that
\be
F' = e^{-4z_n-2au/3+4y} (C + 4 K_*^{-1} z_n') \ .
\ee
For large $u$ (near the horizon), we must have $F' \rightarrow 0$ in order
to satisfy (\ref{asympt}).  This along with the explicit form of $z_n$ (\ref{nexord}) 
fixes the integration constant to be
$C = 2a / K_*$.  Hence
\be
F' = - e^{-4au} \left( \frac{2a}{3 (1-e^{-4au})} \right)^{2/3} \ ,
\label{eFdiff}
\ee   
and thus
\be
F =  \frac{3}{4a^{1/3}} \left(\frac{2}{3} \right)^{2/3} 
\left( 1 - (1-e^{-4au})^{1/3} \right) \ .
\ee
As required by $K(u) = K_* + P^2 F(u)$ (\ref{require}), this
expression satisfies $F(u\rightarrow \infty) \equiv F_* = 0$.  In
other words, the fractional D-string charge $K(u) \rightarrow K_*$ in
the large $u$ limit, as desired.  Notice that $F'<0$ so that, as
advertised in Section 5, $K(u)$ decreases as $u$ increases. Moreover, if $P^2
K_*^{-1} a^{-1/3} \ll 1$, the perturbation caused by $F(u)$ is small
for all values of $u$.  Even at small $u$, $F$ is well behaved
\be
F(u) = 
 \frac{3}{4a^{1/3}} \left( \frac{2}{3} \right)^{2/3} 
 - \frac{6^{1/3}}{2} u^{1/3} 
+ O(u^{4/3})  \ .
\label{smFu}
\ee
In addition, from these small $u$ asymptotics, we find that
at small $u$, we have recovered the extremal fractional D-string
solution (\ref{extone})!

Thus already at the leading order this perturbation theory produces a
solution with the correct extremal fractional D-string asymptotics.
This remarkable fact strengthens our confidence that an exact solution
interpolating between the extremal fractional D-string at small $u$
and the regular non-extremal D-string horizon at large $u$ indeed
exists.  Our perturbed solution should be a good approximation to it
provided that $P^2 K_*^{-1} \ll a^{1/3}$.  This limit corresponds to
high Hawking temperatures, as we now show by matching (\ref{smFu})
with (\ref{smallu}) for small $u$.  We find that
\be
\frac{3 \cdot 6^{1/3}}{8} L_{\rm P}^{-2} = \frac{K_*}{P^2} + 
\left( \frac{2}{3} \right)^{2/3} \frac{3}{4a^{1/3}} \ . 
\label{lpdef}
\ee
On the other hand, the Hawking temperature is determined in terms of the
non-extremality $a$ and the charge near the horizon $K_*$ by the usual
D-string formula (\ref{th})
\be
T = \frac{3}{\pi} \left( \frac{3}{2} \right)^{1/6} a^{1/3} K_*^{-1/2} \ .
\label{tha}
\ee
Comparing with the extremal fractional D-string solution, we expect that 
the critical temperature roughly corresponds to the value of $a$ for which the
naked singularity of the extremal fractional D-string solution and the 
horizon of the regular non-extremal fractional D-string coincide.  
In other words, $a^{1/3} \sim L_{\rm P}^2$.  
Thus, the perturbation theory will correspond to high Hawking
temperatures so long as $a^{1/3} \gg L_{\rm P}^2$.  Comparing
with (\ref{lpdef}), we recover the naive inequality needed for 
our perturbation theory to be valid, namely $P^2 / K_* \ll a^{1/3}$.
To summarize, in our perturbative regime, $T \gg T_c$. 

\subsection{Solutions for other fields}

We begin by solving for the correction to the dilaton
$\phi$.  Using (\ref{eFdiff}) and (\ref{nexord}),
the equation for the dilaton correction (\ref{phi}) becomes
\be
\phi'' = \frac{1}{K_*} \left( \frac{2a}{3} \right)^{5/3}
\frac{e^{-4au}}{(1-e^{-4au})^{2/3}} \ .
\label{phis}
\ee
Integrating once, and keeping in mind the boundary condition 
$\phi' \rightarrow 0$ as $u \rightarrow \infty$, we find
\be
\phi' = -\frac{2a}{3K_*} F = \frac{3}{4aK_*} \left( \frac{2a}{3} \right)^{5/3}
\left( (1 - e^{-4au} )^{1/3} -1 \right)
\ee
Hence, $\phi$ is a decreasing function of $u$.  In other words, the string
coupling decreases as we approach the horizon.
To solve for $\phi$, we may integrate once more:
\be 
\phi = \frac{1}{8 k_*} \left[
-3v^{1/3} + \frac{3}{2} \log \frac{1-v}{1-v^{1/3}} + 
\sqrt{3} \tan^{-1} \frac{2v^{1/3}+1}{\sqrt{3}} - \frac{ \pi \sqrt{3}}{6}
\right]
\label{phie}
\ee
{}From (\ref{bcsmall}), we have fixed the boundary condition such that
$\phi=0$ at $u=0$.  Moreover, to make the formula neater, we have
introduced 
\be
v=1-e^{-4au} \   
\ee
and 
\be
\frac{1}{k_*} \equiv  \frac{1}{K_* a^{1/3}} \left(\frac{2}{3} \right)^{2/3} \ .
\ee

To show that $\phi$ is well behaved at both small $u$ and large $u$,
we can write down some limiting expressions
\be
u \rightarrow \infty \ : \ \ \ \ \phi = \phi_* 
+ \frac{1}{24k_*} (1-v) + O[(1-v)^2] \ , 
\label{phil}
\ee
\be
u \rightarrow 0 \ : \ \ \ \ \phi = -\frac{1}{8k_*} v + \frac{3}{32k_*} v^{4/3} +
+ O(v^2) \ ,
\ee  
where  
\be
\phi_* = \frac{1}{8k_*} \left( -3 + \frac{\pi \sqrt{3}}{6} + \frac{3}{2}\log 3
\right) \ .
\label{pst}
\ee

The relations describing the corrections to the other fields  
(\ref{xidiff})--(\ref{omegaii}), (\ref{etadiff}) are all ordinary, second order,
linear differential equations.  There exists a powerful technique,
called the Lagrange method of variation of parameters, which 
is useful for dealing with this type of ODE.  In particular, given
an ODE of the form
\be
\frac{d^2y}{dx^2} + p(x) \frac{dy}{dx} + q(x) y = g(x) \ , 
\label{ptype}
\ee   
and
two linearly independent solutions $y_1$ and $y_2$ to the corresponding 
homogenous equation $g(x)=0$, we can construct a general solution
corresponding to the case $g(x) \neq 0$:
\be
Y = -Y_1 \int \frac{Y_2 g}{W} dx + Y_2 \int \frac{Y_1 g}{W} dx
+ c_1 Y_1 + c_2 Y_2 \ .
\label{Lsol}
\ee
where $W=Y_1 \frac{dY_2}{dx} - \frac{dY_1}{dx} Y_2$ is the Wronskian.
(Linear independence corresponds to the fact that $W \neq 0$.)

As a first step, we cast the differential
equations for $\omega_i$, $\xi$, and $\eta$, 
(\ref{xidiff})--(\ref{omegaii})
and (\ref{etadiff}), into the form (\ref{ptype}), i.e.
\be
\omega_1'' - \frac{64a^2}{9} \frac{e^{-4au}}{(1-e^{-4au})^2} \omega_1 =
\frac{2a^2}{21k_*} \frac{e^{-4au}}{(1-e^{-4au})^{2/3}}
\ ,
\label{oo}
\ee
\be
\omega_2'' + \frac{32a^2}{9} \frac{e^{-4au}}{(1-e^{-4au})^2} \omega_2 =
-\frac{a^2}{3k_*} \frac{e^{-4au}}{(1-e^{-4au})^{2/3}}
\ ,
\label{ot}
\ee
\be
\xi'' - \frac{32 a^2 e^{-4au}}{(1-e^{-4au})^2} \xi =
-\frac{2a^2}{21k_*} \frac{e^{-4au}}{(1-e^{-4au})^{2/3}}
\ ,
\label{xit}
\ee   
\be
\eta'' - \frac{32a^2 e^{-4au}}{(1-e^{-4au})^2} \eta =
\frac{a^2}{2k_*} \left[
\frac{12 e^{-4au}}{(1-e^{-4au})^2} - \frac{11e^{-4au}}{(1-e^{-4au})^{5/3}} +
\frac{e^{-8au}}{(1-e^{-4au})^{5/3}}
\right] \ .
\label{etat}
\ee
To analyze these differential equations, we again introduce the new radial 
variable $v=1-e^{-4au}$.
Then, (\ref{oo})--(\ref{etat}) become
\be
\ddot{\omega}_1 - \frac{\dot{\omega}_1}{1-v} - \frac{4/9}{v^2(1-v)}\omega_1 = 
\frac{1}{168k_*} \frac{1}{v^{2/3} (1-v)} \ ,
\label{ovo}
\ee
\be
\ddot{\omega}_2 - \frac{\dot{\omega}_2}{1-v} + \frac{2/9}{v^2(1-v)}\omega_2 = 
-\frac{1}{48k_*} \frac{1}{v^{2/3} (1-v)} \ ,
\label{ovt}
\ee
\be
\ddot{\xi} - \frac{\dot{\xi}}{1-v} - \frac{2}{v^2(1-v)}\xi = 
-\frac{1}{168k_*} \frac{1}{v^{2/3} (1-v)} \ ,
\ee
\be
\ddot{\eta} - \frac{\dot{\eta}}{1-v} - \frac{2}{v^2(1-v)}\eta = 
\frac{1}{32k_*} \left[
\frac{12}{v^2(1-v)} - \frac{11}{v^{5/3}(1-v)} + \frac{1}{v^{5/3}}
\right] \ ,
\ee
where the dots denote $d/dv$.  Now, the homogenous equation
\be
\ddot{f} - \frac{\dot{f}}{1-v} - \frac{A}{v^2(1-v)} f = 0
\ee
is solved for generic $A$ by $f(v) = v^\nu \, {}_2F_1(\nu,\nu;2\nu;v)$, where
${}_2F_1$ is the hypergeometric function and $\nu(\nu-1)=A$.  As it
happens, $A=2$ is a degenerate case where the solutions to the homogenous
equation are elementary functions of $v$ (namely, $\frac{1}{v}-\frac{1}{2}$
and $-2+\frac{v-2}{v}\log(1-v)$).  Perhaps surprisingly, we have been
able to find analytic expressions for $\xi$ and $\eta$ using these
homogenous solutions.  Let us start with $\xi$.  In particular
\[
\xi = \frac{1}{56k_* v}
\left[
3v^{1/3}(v-4) + (v-2)\log \frac{(1-v^{1/3})^3}{1-v} 
+2 \sqrt{3} (2-v) \tan^{-1} \frac{2v^{1/3}+1}{\sqrt{3}} 
\right] +
\]
\be
+ b_1 \left(\frac{1}{v} - \frac{1}{2} \right)
+ b_2 \left( -2 + \frac{v-2}{v} \log(1-v) \right) \ .
\label{xie}
\ee
The boundary conditions for $\xi$ are that $\xi = 0$ at 
$v=0$ (\ref{requirem})
and $d \xi / du \rightarrow 0$ as $u \rightarrow \infty$
(\ref{asympt}).  This second boundary condition, in terms
of the $v$ variable, means that $d\xi /dv$ must be finite or zero
in the limit $v\rightarrow 1$ because $dv/du = 4a (1-v)$.
As a result, we find that 
\be
b_1 = -\frac{\pi \sqrt{3}}{84k_*} \ ; \ \ \ \ 
b_2 = - \frac{1}{28k_*} \ .
\ee

Next, we solve for $\eta$:
\[
\eta = \frac{3}{64k_*v} \left[
6v^{1/3}(v-2) - 8 + (v-2)\log \frac{(1-v^{1/3})^3}{1-v} 
+2 \sqrt{3} (2-v) \tan^{-1} \frac{2v^{1/3}+1}{\sqrt{3}}
\right]
\]
\be
+ c_1 \left(\frac{1}{v} - \frac{1}{2} \right)
+ c_2 \left( -2 + \frac{v-2}{v} \log(1-v) \right) \ .
\label{etae}
\ee
Recall at the end of Section 5, we expected that the correction
to the original dilaton $\Phi$ should decrease as $u\rightarrow \infty$,
reflecting the corresponding decrease in $K$.  Indeed, from
(\ref{phie}) and (\ref{etae}), it is clear that this correction
$3\phi/2 - 2\eta$ (see (\ref{zptrans})) is indeed a decreasing function of $u$.
The boundary conditions for $\eta$ are again that 
$d \eta / du \rightarrow 0$ as $u \rightarrow \infty$
(\ref{asympt}).  However the small $u$ boundary condition
is merely the fact that $\eta$
does not diverge at small $u$.  The integration constants which 
satisfy these conditions are
\be
c_1 = \frac{1}{32k_*} (12 - \pi \sqrt{3}) \ ; \ \ \ \
c_2 = -\frac{3}{32k_*} \ .
\ee

To show that $\xi$ and $\eta$ are well behaved at both large $u$ and
small $u$, we look at the asymptotics.  First, we examine the long
distance asymptotics ($v\rightarrow 0$ or equivalently $u \rightarrow
0$) of $\xi$ and $\eta$:
\be
\xi = \frac{3}{784 k_*} v^{4/3} - \frac{1}{168k_*} v^2 + O(v^{7/3}) \ ,
\ee
\be
\eta = -\frac{3}{16k_*}  + \frac{9}{64k_*} v^{1/3} + 
\frac{9}{896k_*} v^{4/3} - \frac{1}{64k_*}v^2 + O(v^{7/3}) \ .
\ee
The first and second terms in the $\eta$ expansion agree with the 
extremal fractional D-string solution, as one can see by comparing
with (\ref{lpdef}) and (\ref{smallu}).  

Next we consider the expansions of $\xi$ and $\eta$ near the horizon,
$u\rightarrow \infty$ or equivalently $v=1$:
\be
\xi = \xi_* + \left(2\xi_* - \frac{1}{168 k_* } \right) (1-v)
+ O[(1-v)^2] \ ,
\label{xil}
\ee
\be
\eta = \eta_* + \left( 2\eta_* +\frac{1}{32k_*} \right)(1-v) + O[(1-v)^2] \ .
\label{etal}
\ee
The horizon values $\eta_*$ and $\xi_*$ are
\be
\xi_* = \frac{1}{168k_*} (-15+ \pi \sqrt{3} + 9 \log 3 ) \ ,
\label{xst}
\ee
\be
\eta_* = \frac{1}{64k_*} (-18 + \pi \sqrt{3} + 9 \log 3) \ .
\label{est}
\ee

We were careful to calculate $\xi_*$, $\eta_*$ and
$\phi_*$ because they show up as corrections to
observables such as the Hawking temperature (\ref{th}) and the horizon 
area.  On the other hand, the volume of the compact space, this squashed
$Q^{1,1,1}$, does not depend on the 
squashing factors $\omega_1$ and $\omega_2$; the squashing
factors cancel from the observables we are interested in 
calculating.  As a result, and because the homogenous solutions
to (\ref{ovo}) and (\ref{ovt}) are more
complicated, we shall be more cavalier
in our treatment of the asymptotics for $\omega_1$ and $\omega_2$.
At small $u$, (\ref{oo}) and (\ref{ot}) become 
\be
\omega_1 = p_1 u^{4/3} + \frac{a^{4/3} 4^{1/3}}{70k_*} u^{4/3} \log u
+ O(u^{7/3}) \ ,
\ee
\be
\omega_2 = p_2 u^{1/3} + p_3 u^{2/3} -\frac{a^{4/3} 2^{2/3}}{8k_*} u^{4/3}
+ O(u^{7/3}) \ .
\ee
The $p_i$, $i=1,2,3$, are undetermined constants of integration
which can be determined numerically, keeping in mind the boundary conditions
$w_i(0)=0$ and $w_i'(\infty)=0$.  The large $u$ asymptotics are
\be
\omega_1 = \omega_{1*} + \left(\frac{4}{9}\omega_{1*} +
\frac{1}{168 k_*} \right) e^{-4au} + O(e^{-8au}) \ ,
\ee
\be
\omega_2 = \omega_{2*} + \left(-\frac{2}{9} \omega_{2*} -
\frac{1}{48 k_*} \right) e^{-4au} + O(e^{-8au}) \ .
\ee

We can now go a bit further than was done in the previous sections and
determine the order $P^2$ corrections to the temperature and 
the entropy.  The metric (\ref{metric}) can be recast into the form
\[
ds_{10E}^2 = \left( \frac{4a}{K_*} \right)^{3/4} e^{3P^2(\eta + \phi/4)}
v^{-3/4} \left[ (1-v)dX_0^2 + dX_1^2 \right] +
\]
\be 
\left( \frac{4a}{K_*} \right)^{-1/4} \left( \frac{2a}{3} \right)^{1/3} 
v^{-1/12} 
\left( \frac{1}{36} e^{P^2(14\xi-\eta-\phi/4)} \frac{dv^2}{v^2(1-v)} +
e^{P^2(2\xi-\eta-\phi/4)} (dM_7)^2 \right) \ .
\ee
Using the large $u$ asymptotics for $\xi$ (\ref{xil}), $\eta$ (\ref{etal}), 
and $\phi$ (\ref{phil}), we obtain an explicit expression for the entropy
per unit volume divided by the temperature squared:
\be
\frac{S}{VT^2} = \alpha K_*^{3/2} e^{3P^2(7\xi_* - 2\eta_* - \phi_*/2)} =
\alpha K_*^{3/2} \left( 1 + \frac{3P^2}{8k_*} + O(P^4/k_*^2) \right) \ ,
\label{dsqbn}
\ee 
where $\alpha$ is a constant of order unity.  In the last equality,
we have used the values for $\xi_*$ (\ref{xst}), 
$\eta_*$ (\ref{est}), and $\phi_*$ (\ref{pst}).  
Note that the transcendental numbers $\log 3$ and $\pi$ drop out of the
linear combination of $\xi_*$, $\phi_*$, and $\eta_*$.

Using (\ref{tha}) and (\ref{lpdef}), one finds that
\be
K_* \sim \frac{P^2}{L_{\rm P}^2} \left[ 1 - 
\frac{2^{5/2} L_{\rm P}^3}{\pi 3^{1/2} PT} +\ldots \right] \ .
\ee
Hence, from (\ref{dsqbn}),
\be
\frac{S}{VT^2} \sim \frac{P^3}{L_{\rm P}^3} \left[ 1 - 
\frac{2^{5/2} L_{\rm P}^3}{\pi 3^{1/2} PT} +\ldots \right] \ .
\ee
As expected, we find that the entropy ratio $S/VT^2$ increases toward
a limiting value as $T$ increases.  The important point is that both
the number of D-strings at the horizon $K_*$ and the entropy ratio
$S/VT^2$ depend in the same way on the temperature.  This picture is
consistent with gauge theory where the number of D-strings should
correspond roughly to the number of degrees of freedom in the theory.
As the number of D-strings grows, so should the entropy.

\section{Remarks}

We have presented, within the framework of perturbation theory in the number
of fractional D-strings, a well behaved non-extremal fractional D-string
solution.  This finite temperature solution breaks supersymmetry but
preserves the $U(1)$ symmetry of the transverse conical space.  It would be good if an additional KS type solution could be
found for these fractional D-strings in a $Q^{1,1,1}$ background.  In
analogy with the KS solution for fractional D3-branes, we would expect the
corresponding solution to involve blowing up the tip of the cone
over $Q^{1,1,1}$ in a way that keeps the five-cycle finite but allows the
two cycle to become vanishingly small. In addition, we expect this solution
would preserve supersymmetry but would break the $U(1)$ symmetry of the
$Q^{1,1,1}$.  If such a KS type solution is found, then we can use the
finite temperature solution found here as evidence for a chiral symmetry
breaking phase transition.  We leave construction of such a deformation for
future work.

Another interesting direction to pursue is construction of a finite
temperature solution for the fractional D2-branes.  In \cite{Cvetic3}, two
examples of a KS type deformation of the fractional D2-branes were found.
Each example involves a different six dimensional Einstein
manifold -- an $S^2$ bundle over either $S^4$ or ${\bf CP}^4$.  It should
be straightforward to introduce appropriate squashing functions to the cones
over these two Einstein spaces, thereby producing an ansatz which admits
well behaved fractional D2-branes at finite temperature, at least in
perturbation theory.

{}Finally, it would be extremely interesting to find a way to use the
systems of differential equations found here \eqsysf\ and in
\cite{BHKPT} to see what happens close to the expected phase
transition.  Unfortunately, to date we know of no such analytic
solutions.  Moreover, there are many integration constants involved,
some of which are set in the IR, some in the UV, which makes any kind
of numerical shooting algorithm a tedious prospect. Still, the authors
of the current paper were mildly surprised that the set of
differential equations \eqsysf\ proved tractable in the first order in
perturbation theory, and it seems likely that there are some more
surprises waiting.

\section*{Acknowledgments}
We are grateful to I.~R.~Klebanov and M.~Rangamani
for many useful discussions.  The
work of C.~P.~H. was supported in part by the DoD.  P.~O. is supported
in part by an NSF Graduate Research Fellowship.

\end{document}